\documentclass[conference]{IEEEtran}
\IEEEoverridecommandlockouts
\usepackage{cite}
\usepackage{amsmath,amssymb,amsfonts}
\usepackage{algorithmic}
\usepackage{graphicx}
\usepackage{listings}
\usepackage{textcomp}
\usepackage{xcolor}
\graphicspath{{images/}} 
\usepackage{comment}

\def\BibTeX{{\rm B\kern-.05em{\sc i\kern-.025em b}\kern-.08em
    T\kern-.1667em\lower.7ex\hbox{E}\kern-.125emX}}
\begin{document}

\makeatletter
\newcommand{\linebreakand}{%
\end{@IEEEauthorhalign}
\hfill\mbox{}\par
\mbox{}\hfill\begin{@IEEEauthorhalign}
}
\makeatother

\title{Augmented Knowledge Graph Querying leveraging LLMs}

\author{
	\IEEEauthorblockN{Marco Arazzi}
	\IEEEauthorblockA{Department of Electrical, Computer\\ and Biomedical Engineering,\\ University of Pavia, Italy} \hspace{1cm} 
	\and
	\IEEEauthorblockN{Davide Ligari}
	\IEEEauthorblockA{Department of Electrical, Computer\\ and Biomedical Engineering,\\ University of Pavia, Italy}\\[1em]
	\linebreakand
	\IEEEauthorblockN{Serena Nicolazzo}
	\IEEEauthorblockA{Department of Computer Science,\\ University of Milan, Italy} 
	\and
	\IEEEauthorblockN{Antonino Nocera}
	\IEEEauthorblockA{Department of Electrical, Computer\\ and Biomedical Engineering,\\ University of Pavia, Italy}
}
\maketitle

\begin{abstract}
    
    Adopting Knowledge Graphs (KGs) as a structured, semantic-oriented, data representation model has significantly improved data integration, reasoning, and querying capabilities across different domains. This is especially true in modern scenarios such as Industry 5.0, in which the integration of data produced by humans, smart devices, and production processes plays a crucial role. However, the management, retrieval, and visualization of data from a KG using formal query languages can be difficult for non-expert users due to their technical complexity, thus limiting their usage inside industrial environments. For this reason, we introduce SparqLLM, a framework that utilizes a Retrieval-Augmented Generation (RAG) solution, to enhance the querying of Knowledge Graphs (KGs). SparqLLM executes the Extract, Transform, and Load (ETL) pipeline to construct KGs from raw data. It also features a natural language interface powered by Large Language Models (LLMs) to enable automatic SPARQL query generation. By integrating template-based methods as retrieved-context for the LLM, SparqLLM enhances query reliability and reduces semantic errors, ensuring more accurate and efficient KG interactions. Moreover, to improve usability, the system incorporates a dynamic visualization dashboard that adapts to the structure of the retrieved data, presenting the query results in an intuitive format. Rigorous experimental evaluations demonstrate that SparqLLM achieves high query accuracy, improved robustness, and user-friendly interaction with KGs, establishing it as a scalable solution to access semantic data.
\end{abstract}


\section{Introduction}

In recent years, the diffusion of digital technologies has led to exponential growth in the amount of data generated and stored worldwide.
This is even more evident in the case of Industry 5.0, which shifts toward intelligent, human-centric, and eco-friendly industrial ecosystems. In this novel industrial paradigm, the Internet of Things (IoT) and its evolution, the Internet of Everything (IoE), play a crucial role by supporting the interaction between devices, data, processes, and people, through a highly interconnected environment.
The exponential growth of produced data that such an environment can bring, combined with data diversity and heterogeneity, has made data management and integration increasingly complex and prone to errors.

Traditional data storage and processing techniques often fall short when dealing with large-scale, unstructured,
and multi-source datasets. In addition, the growing demand for real-time insight and decision-making based on
integrated data further highlights the limitations of conventional approaches.
Addressing these challenges requires sophisticated methodologies and tools capable of handling the entire data lifecycle of integration and querying.

One promising solution is using \textit{Knowledge Graphs} (KGs, hereafter), which leverage semantic technologies to represent and
interconnect data meaningfully. By providing a structured and machine-interpretable representation of information,
KGs facilitate seamless data integration, interoperability, and accessibility \cite{Arazzi_2024}. 
However, retrieving dynamic data from such complex models requires expertise in formal query languages, like SPARQL \cite{hogan2020sparql}, thus heavily limiting their usability for non-technical employees.

The integration of KGs with advanced tools such as Large Language Models (LLMs, hereafter) opens new avenues for Natural
Language Processing and intelligent data querying. By enabling intuitive and human-friendly interfaces,
these technologies fill the gap between raw data and actionable knowledge, allowing users to interact with datasets more effectively~\cite{arxiv_llms_nlp}.
However, the use of KGs in this context presents additional challenges. Existing systems often do not
generate precise and efficient SPARQL queries \cite{hogan2020sparql}. Another critical aspect when attempting to integrate advanced AI tools with semantic technologies is that such systems are often limited by query hallucination, semantic misalignment, and lack of adaptability to different data sets. Ultimately, this may lead to answers that fall short of user expectations.
Moreover, as an additional key challenge, the results of a produced query must be transformed into intuitive, human-readable outputs, without requiring the intervention of technical experts and data analysts.

This paper tackles the above issues by developing a comprehensive system, called SparqLLM, that integrates semantic technologies with state-of-the-art AI tools. SparqLLM is designed as a Retrieval-Augmented Generation (RAG) framework that automates the generation of SPARQL queries from natural language questions while producing the most adequate data visualizations to return the obtained results. By leveraging template-based methods as retrieved-context for the LLM, SparqLLM enhances query reliability and minimizes semantic errors. Our goal is to improve the accuracy, usability, and reliability of Knowledge Graphs (KGs), enabling more intuitive and effective interaction with semantic data. An extensive evaluation confirmed the system's effectiveness, demonstrating high accuracy in query generation and strong visualization capabilities. The results highlight the potential of combining semantic technologies with advanced AI to simplify access to KGs and improve their usability. This work contributes to bridging the gap between raw data and actionable insights by providing a scalable, user-friendly framework for semantic data management and analysis.

The outline of this paper is as follows. In Section \ref{sec:related}, we discuss the related papers in the state of the art. In Section \ref{sec:kg} we describe the KG used in this work and its generation. Section \ref{sec:approach} gives details of the proposed approach for KG interrogation. In Section \ref{sec:experiments}, we present the experiments carried out to test our approach and show its performance. Finally, Section \ref{sec:conclusion} examines intriguing leads as future work and draws our conclusions.

\section{Related Work}
\label{sec:related}

Natural Language Interfaces (NLIs) are essential for facilitating access to KGs, enabling users to interact with these complex data structures through conversational natural language \cite{liang2021querying}. By translating unstructured input into formal query languages such as SPARQL, NLIs bridge the gap
between user-friendly interaction and the highly structured nature of semantic data. The development of NLIs addresses a significant challenge
in making RDF-based knowledge graphs more accessible to non-technical
users, thereby expanding their applicability across diverse fields such as healthcare, e-commerce, and education.

LLMs have recently shown great potential in the
translation of natural language into SPARQL queries. By leveraging their
ability to process and generate complex text, LLMs offer a robust framework for automating query generation, reducing the need for manual intervention, and making knowledge graphs accessible to non-expert users \cite{Pan_2024}.
Early systems like SGPT leverage transformer-based architectures to generate SPARQL queries, addressing structural and semantic requirements~\cite{Rony2022SGPT}.
Another significant approach involves fine-tuning the OpenLLaMA model for SPARQL query generation.
By adapting this LLM to specific datasets containing question-to-SPARQL pairs, the model effectively translates natural language questions into SPARQL
queries over life science knowledge graphs. Despite its success, the model's performance is contingent on the quality and representativeness of the training data,
which may limit its generalizability to other domains~\cite{Reyes2024SPARQLGA}. BART, another transformer-based model, has also been applied to SPARQL query generation.
Its encoder-decoder structure is well-suited for handling long contexts and generating queries that require nested reasoning.
For instance, the NLQxform system employs a fine-tuned BART model to translate natural language questions into SPARQL,
integrating steps like entity linking and template-based corrections to improve accuracy. While BART performs well in structured domains,
it faces challenges when dealing with noisy or incomplete inputs~\cite{wang2023nlqxformlanguagemodelbasedquestion}.

Template-based approaches complement these methods by providing deterministic frameworks for query generation. Systems like CatSQL and BERT-based methods enhance
SQL generation by combining predefined templates with semantic corrections and contextual embeddings.

Although these approaches ensure interpretability and efficiency, their reliance on predefined patterns limits flexibility in handling complex or ambiguous queries~\cite{Fu2023CatSQL, Long2021Bert-based}. 
Despite advances in data integration, ETL, and query generation, challenges remain in scaling these systems for diverse domains and ensuring seamless
interaction with KGs. This work aims to integrate semantic technologies with AI-driven methods for efficient and user-friendly data querying and visualization.

\section{Knowledge Graph Construction}
\label{sec:kg}
In this section, we detail the structure of the KG developed for this work, including its basic building blocks and the ontological
schema. Additionally, we outline the implementation of the ETL pipeline for the KG construction.

\subsection{Knowledge Graph Ontology}

The development of semantic systems capable of handling complex and interconnected data relies heavily on KGs and their underlying ontologies.
In this work, a KG is designed to integrate information about Industry 5.0 by focusing on an existing ontology, specifically tailored for this domain, named the SemIoE (Semantic Internet of Everything) \cite{Arazzi_2024}.
In particular, the SemIoE ontology extends the Semantic Sensor Network (SOSA/SSN) \cite{compton2012ssn}, the Building Topology Ontology
(BOT) \cite{world2014organization}, and the Organization Ontology (ORG) \cite{rasmussen2021bot}, to model interactions in the Internet of Everything (IoE) within Industry 5.0 ecosystems.
It incorporates concepts such as \textit{ioe:Agent} and \textit{ioe:SmartObject}, which represent overarching systems managing multiple components,
thereby addressing the dynamic needs of IoE environments.


The constructed graph is thus grounded in established ontologies, thus achieving semantic precision while supporting scalability and interoperability.
It enables advanced querying and reasoning, laying a solid foundation for applications requiring integrated and contextually rich data representations.

\subsection{ETL Pipeline for Knowledge Graph Construction}

The Extract, Transform, Load (ETL) pipeline serves as the backbone for transforming raw data into a structured, semantically enriched KG. 
To prepare data from an heterogeneous IoE for integration into a KG, a rigorous data-cleaning process is typically required to ensure consistency, standardization, and semantic integrity. Key tasks may include \textit{(i) Standardizing Formats} in which temporal, numerical, and textual data from different data sources (smart objects, organization-specific information, and so forth) are reformatted to comply with established conventions, enhancing uniformity and compatibility;  \textit{(ii) Aligning with Ontology}, in which data is categorized and structured to align with the ontology, grouping related entities for semantic coherence. After this step, the cleaned data can be transformed into the RDF format using the RDF Mapping Language (RML), a versatile tool for integrating heterogeneous datasets. RML mapping rules define how the source data is translated into RDF triples, specifying:
\textit{(i) the Subjects}, that are the resources described in the RDF graph and are generated dynamically using URI templates; 
\textit{(ii) the Predicates}, that are the properties or relationships linking subjects to values or other resources; and, finally, \textit{(iii) the Objects}, that are the values or references completing the triple structure. RML mapping rules ensure semantic alignment with the reference ontology. In our implementation, we used the RML.io Java implementation for its reliability and performance, generating RDF triples ready for ingestion.

For data ingestion, the RDF triples can be uploaded into an RDF triplestores. In our solution, we make explicit reference to GraphDB, a scalable and high-performance RDF database. GraphDB's REST API facilitated seamless programmatic interaction for uploading RDF data, executing SPARQL queries, and managing updates. Database configurations, including reasoning and indexing, can be adjusted to enhance query performance and enable the inference of new relationships based on the ontology. This systematic approach can ensure the efficient storage and retrieval of semantically enriched data, supporting complex KG operations.

\section{Knowledge Graph Querying}
\label{sec:approach}

This section presents the design and implementation of the RAG component of SparqLLM, which transforms natural language queries into SPARQL queries, facilitating interaction with a KG. 

\begin{figure*}[ht]
    \centering
    \includegraphics[width=\textwidth]{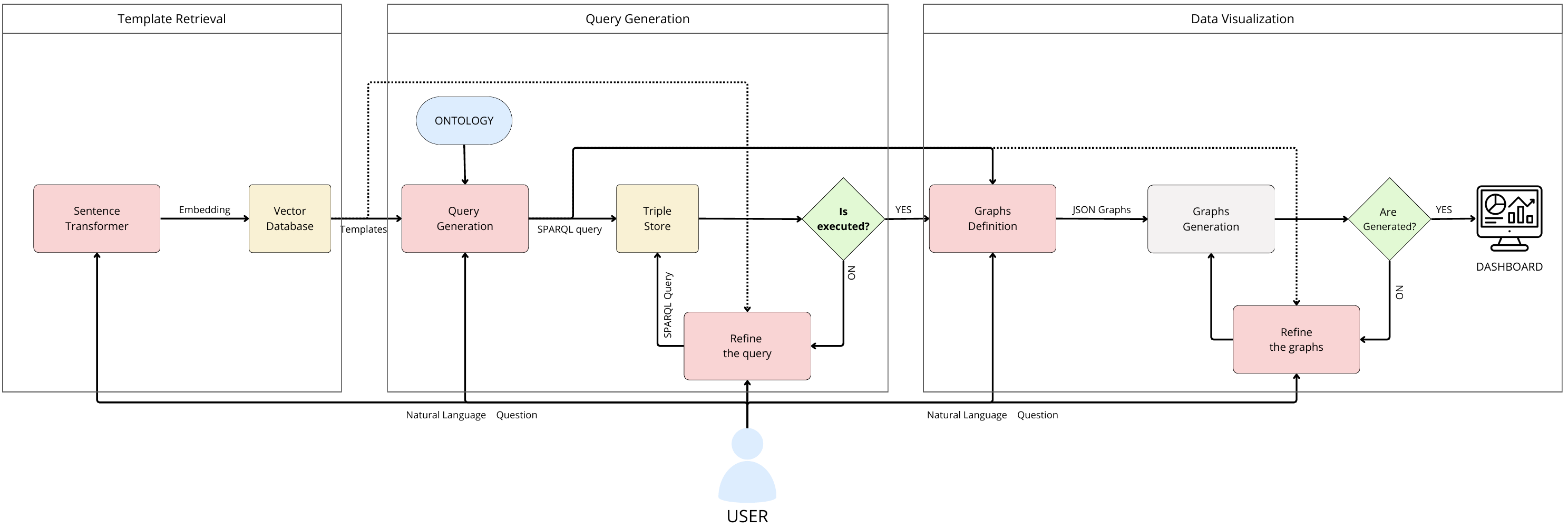}
    \caption{Visual representation of system's architecture }\label{fig:overall_system}
\end{figure*}
As shown in Figure \ref{fig:overall_system}, this part of the system consists of three main functionalities, namely: 

\begin{itemize}
    \item \textbf{Template Retrieval}: in which the system identifies the most relevant SPARQL query templates based on semantic similarity to the user's question.
    \item \textbf{Query Generation}: the retrieved templates, combined with the user's query and the KG ontology, are processed by the LLM to generate a structured SPARQL query. In this step, the generated query is executed on the KG. 
    \item \textbf{Data Visualization}: the query results are transformed into visual representations.
\end{itemize}
    
In the following subsections, the implementation details of each component are presented.

\subsection{Template Retrieval}

Templates act as predefined parametric analysis types that must be elaborated by the engine. These templates enable the transformation of user intents into structured and precise queries.
Instead of generating SPARQL queries from scratch, the system retrieves and adapts an appropriate template. This approach reduces the complexity of query generation, as the language model (LLM) focuses on refining an existing structure rather than creating a new one. Templates, therefore, serve as a foundation that minimizes errors and accelerates query construction. Templates also ensure semantic alignment, because they capture a wide range of query types and user intents. By leveraging these templates, the system can map the semantic meaning of a natural language query to a structured query format. This alignment bridges the gap between user-friendly natural language and the technical structure of SPARQL queries. Each template represents a specific type of query, such as:

\begin{itemize}
    \item \textit{SELECT:} queries that retrieve specific attributes of entities without performing aggregation.
    \item \textit{GROUP BY:} queries involving aggregation, such as averages or counts.
    \item \textit{FILTER:} queries requesting information filtered by certain attributes, such as date ranges or entity properties.
\end{itemize}

A set of 15 templates was created for each query type and entity in the KG, leading to a total of 360 templates. After this, the templates are integrated into the vector database. Indeed, as typically done on traditional RAG systems, a vector database was used to efficiently store and retrieve SPARQL templates. Unlike traditional databases, this kind of database is better suited for managing high-dimensional embeddings and performing similarity searches. Each SPARQL template was transformed into a high-dimensional vector embedding using an embedding model. In case of complex templates, the embeddings of a template can be enriched with natural language descriptions to improve retrieval accuracy.

Moreover, efficient retrieval was enabled through indexing. The \texttt{IVF\_FLAT} indexing method was used, which organizes vectors into clusters and allows for faster search within the most relevant clusters. This approach reduces the computational cost of similarity searches while maintaining accuracy.

At this point several embedding models can be used to match templates to user queries.

Embedding models play a key role in matching user queries to templates. In this work, sentence embeddings were used to capture the semantic meaning of queries. Several embedding models were evaluated based on their retrieval and semantic textual similarity (STS) performance. In addition to performance metrics, model size was a critical factor in the selection process. While larger models typically achieve higher accuracy, they require significant computational resources, limiting their portability and scalability. Consequently, smaller and moderately sized models were prioritized to ensure practicality without compromising performance.
Table~\ref{tab:embeddings} summarizes the evaluated models.

\begin{table}[]
\centering
\caption{Comparison of the sentence embedding models according to MTEB leaderboard \cite{muennighoff2022mteb}.}
\label{tab:embeddings}
\begin{tabular}{|p{2.7cm}|p{1.1cm}|p{1.3cm}|p{1cm}|p{0.7cm}|}
\hline
\textbf{Model} & \textbf{Model Size (M)} & \textbf{Embedding Dimension} & \textbf{Retrieval Score} & \textbf{STS Score} \\ \hline
NV-Embed-v2 \cite{lee2024nvembed} & $7851$ & $4096$ & $62.65$ & $84.31$ \\
bge-en-icl \cite{chen2024bgem3} & $7111$ & $4096$ & $62.16$ & $84.24$\\
stella-base-en-v2 & $55$ & $768$ & $50.1$ & $83.02$\\
mxbai-embed-large-v1 \cite{seanopen} & $335$ & $1024$ & $54.39$ & $85$\\
blade-embed & $335$ & $1024$ & $53.3$ & $85.04$\\
jina-embeddings-v3 \cite{sturua2024} & $572$ & $1024$ & $53.88$ & $85.81$\\
all-roberta-large-v1 & $335$ & $1024$ & $-$ & $-$\\
bilingual-embedding-large \cite{reimers2019sentence} & $560$ & $1024$ & $-$ & $86.02$\\
\hline
\end{tabular}
\end{table}

\subsection{Query Generation}
Generating SPARQL queries from natural language inputs allows users to interact with KGs without requiring technical expertise in structured query languages. By transforming intuitive user queries into structured SPARQL syntax, the system makes querying accessible to non-experts while ensuring precision and accuracy.

This process leverages Large Language Models (LLMs) within a RAG framework, utilizing the previously retrieved templates to craft the LLM prompt. Open-source models from the LLAMA and QWEN families were chosen for this purpose due to their strong performance across various benchmarks and their flexibility in different computational environments. A range of models, varying in size, was evaluated to strike a balance between accuracy and resource efficiency.

The models' capabilities in understanding and reasoning were benchmarked using datasets like ARC-C, GPQA, and Hellaswag. Table~\ref{tab:llama_benchmark} provides an overview of the performance metrics (in terms of accuracy), demonstrating the trade-offs between different model sizes and configurations.

\begin{table}[]
\centering
\caption{Performance of LLAMA and QWEN models on various benchmarks.}
\label{tab:llama_benchmark}
\begin{tabular}{|l|p{0.8cm}|p{1cm}|p{0.8cm}|p{1cm}|}
\hline
\textbf{Benchmark} & \textbf{Llama 3.2 3B} & \textbf{Llama 3.1 70B} & \textbf{Llama 3.1 8B} & \textbf{Qwen 2 72B} \\ \hline
ARC-C & $78.6$ & $94.4$ & $83.4$ & - \\
CPQA & $32.8$ & $39.5$ & $30.4$ & $42.4$\\
Hellaswag & $69.8$ & - & - & - \\
\hline
\end{tabular}
\end{table}

The query generation process, visible in Figure \ref{fig:query_generation}, involves several steps to ensure the system accurately translates user input into SPARQL queries tailored to the KG. The workflow starts by retrieving a relevant query template based on the user's question. This template, along with the user's input and the KG ontology, is provided to the LLM. The ontology offers a complete view of domain-specific entities, relationships, and schema rules, ensuring that the generated query is semantically and structurally correct. A custom-designed prompt guides the model to prioritize templates and ontology-specific terms, reducing ambiguity and helping the model generate concise, valid queries. Once a query is generated, it is executed on the KG. If the query fails, the system enters a feedback loop where the model analyzes the error, refines the query, and retries. This iterative approach continues until a valid query is produced.

\begin{figure}[ht]
    \centering
    \includegraphics[width=\columnwidth]{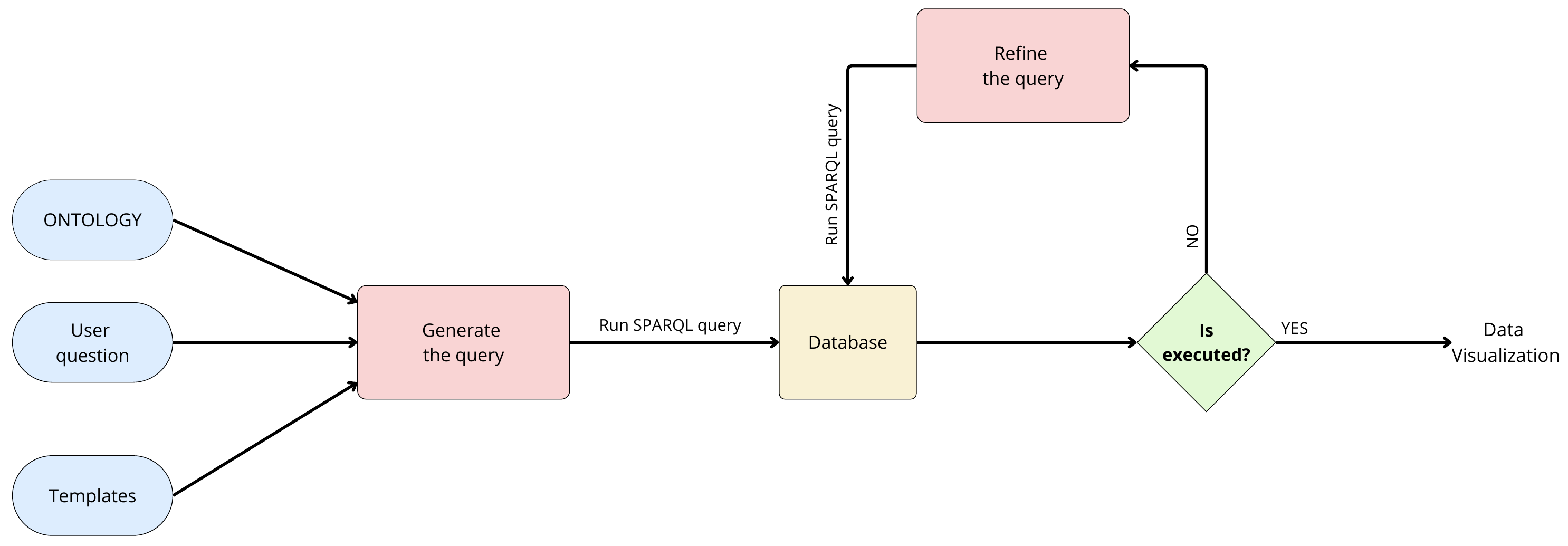}
    \caption{Overview of the SPARQL query generation workflow.}\label{fig:query_generation}
\end{figure}

This multi-step process combines the advantages of predefined templates, ontology guidance, and LLM adaptability. It ensures accurate and meaningful query generation while maintaining user-friendliness. The iterative refinement also enhances reliability, allowing users to retrieve the desired information even in complex scenarios. By making querying intuitive and robust, the system empowers users to access KGs without requiring specialized technical skills.

\subsection{Data Visualization}

SPARQL queries often yield extensive datasets, making it challenging for users to interpret results effectively. To overcome this issue, the system employs advanced data visualization techniques that transform raw query outputs into clear, interactive visual representations. These visualizations allow users to explore data intuitively and analyze results from multiple perspectives. Effective data visualization requires techniques that not only clarify complex data but also enable user interaction. Plotly, a Python library was selected for its ability to create dynamic, customizable visualizations with features like zooming, filtering, and interactive tooltips. This ensures that users can engage with the data at a deeper level. The system incorporates a diverse range of visualizations to accommodate the variety of data structures returned by SPARQL queries, including Bar Charts, Line Charts, Scatter Plots, Violin and Box Plots, Pie Charts, and Tables.
By providing a variety of visualization options, the system ensures flexibility in how data can be analyzed, catering to diverse user needs. The dashboard integrates SPARQL query results into an accessible interface, automatically tailoring visualizations to the data structure and content. A Large Language Model (LLM) assists in this process by selecting appropriate visualization types and generating the corresponding Python code. The visualization workflow is iterative, enabling refinement and error handling at every step.

The process starts with an evaluation of whether the query results are best represented graphically or in a tabular format. For example, queries like ``List all IoT smart objects and their categories'' are suited for tables, whereas data with numerical trends or relationships are better visualized using plots. In the case of graphical representation, the LLM selects a chart type from the predefined list and generates Python code to create the visualization.

\begin{lstlisting}[language=python, caption={Example of Python code generated for visualization.}]
go.Scatter(
    x=df[x],
    y=df[y],
    mode="markers",
)
\end{lstlisting}

The generated code is executed, and if errors arise, the system provides feedback to the LLM, prompting it to refine the code. This cycle continues until the visualization is successfully created. Additionally, the LLM generates descriptive titles and axis labels, enhancing the readability and interpretability of the charts. To ensure accuracy, the LLM is provided with comprehensive contextual inputs at each step. These include
the \textit{User Query} providing context for the overall visualization objective; the \textit{SPARQL Query Structure}, which ensures alignment between the data's schema and the visualization; and the \textit{Data Summaries} including metadata, such as descriptive statistics, assisting the LLM in grasping the data's attributes. By breaking the visualization task into manageable stages, the system ensures robust error handling and precision. This approach also improves monitoring and control, enabling adjustments to be made at each step. The resulting dashboard dynamically adapts to query results, offering clear and engaging visual representations. Whether users are exploring trends, relationships, or distributions, the system ensures the data is presented in the most insightful and user-friendly manner possible.

\section{Experiments}
\label{sec:experiments}

In this section, we present the set of experiments that validate our framework. In particular, we describe the tests that evaluate all the system components, namely the Template Retrieval, the Query Generation, and Data Visualization components.

\subsection{Evaluation of the Template Retrieval Component}

This section presents the experiment that evaluates the Template Retrieval component's ability to retrieve templates with \textit{Target} attributes matching user queries. The evaluation leverages a manually curated dataset and assesses performance across multiple configurations, including embedding models, similarity metrics, embedding types, and the number of templates retrieved. 

The dataset consists of $100$ samples for each combination of query class and entity, ensuring diversity and completeness. It contains six columns: \textit{Question} (the query in natural language), \textit{Query} (the SPARQL query), \textit{Class} (query type, such as \texttt{SELECT}, \texttt{GROUP BY}, or \texttt{FILTER}), \textit{Entity} (knowledge graph entity), \textit{Target} (a combination of \textit{Class} and \textit{Entity}), and other metadata.

The following metrics are employed to quantify system performance and identify areas for improvement:

\begin{itemize}
    \item \textbf{Accuracy} is defined as the proportion of correctly retrieved templates out of the total number of retrieved templates $Accuracy = \frac{TP}{TP+FP}$, where TP (True Positives) is the number of templates retrieved with a Target attribute matching the user's question, whereas FP (False Positives) is the number of templates retrieved with a Target attribute different from the user's question.
    \item \textbf{Matthews Correlation Coefficient (MCC)} is a metric that evaluates the performance of a classification system while considering all elements of the confusion matrix.
    In the multiclass case, it is computed through the following formula $MCC = \frac{\sum_k\sum_l\sum_m C_{kk}C_{lm} - C_{kl}C_{mk}}{\sqrt{\sum_k{T_k}\cdot\sum_k{P_k}}}$, where $T_k = \sum_l C_{kl}$: the total number of templates belonging to class $k$ (true labels), and
    $P_k = \sum_l C_{lk}$: the total number of templates retrieved as class $k$.
\end{itemize}

The first experiment is related to the Embedding Models and Similarity Metrics. The results of these analyses are reported in Table~\ref{tab:embedding_models_comp_similarity_metrics_comp}.
The \texttt{jina-embeddings-v3} model outperformed others, achieving an accuracy of $0.81$ and an MCC of $0.8$.
This indicates its superior capability to correctly retrieve templates and maintain balanced performance across diverse query scenarios. On the other hand, models such as stella-base-en-v2 and bilingual-embedding-large showed relatively lower performance,
with Accuracy values of $0.68$ and $0.69$, respectively.
Among similarity metrics, COSINE and IP demonstrated comparable performance, with IP selected for its computational efficiency. 

\begin{table}[ht]
\centering
\caption{Comparison of embedding models based on Accuracy, MCC, and similarity metrics (COSINE, IP, L2).}
\label{tab:embedding_models_comp_similarity_metrics_comp}
\resizebox{\columnwidth}{!}{%
\begin{tabular}{@{}|l|c|c|c|c|c|@{}}
\hline
\textbf{Model} & \textbf{Accuracy} & \textbf{MCC} & \textbf{COSINE} & \textbf{IP} & \textbf{L2} \\ 
\hline
all-roberta-large-v1 & $0.72$ & $0.70$ & $0.72$ & $0.72$ & $0.72$\\
blade-embed & $0.74$ & $0.73$ & $0.74$ & $0.76$ & $0.74$\\
bilingual-embedding-large & $0.69$ & $0.68$ & $0.69$ & $0.69$ & $0.69$\\
jina-embeddings-v3 & $0.81$ & $0.80$ & $0.81$ & $0.81$ & $0.81$\\
mxbai-embed-large-v1 & $0.73$ & $0.72$ & $0.73$ & $0.73$ & $0.73$\\
stella-base-en-v2 & $0.68$ & $0.66$ & $0.72$ & $0.68$ & $0.72$\\ \hline
\end{tabular}%
}
\end{table}

The second experiment analyzes different types of embedding to be associated with a template. In particular, we considered a direct embedding of the template itself, the embedding of a natural language description associated with it, or a combination of the two. The results are visible in Table~\ref{tab:embedding_types_comp}. From these results, we can conclude that direct embeddings consistently yielded the best results, while description embeddings underperformed due to their general nature. Combined embeddings offered balanced performance but were less precise than direct embeddings. Based on these findings, direct embeddings were adopted for the system.

\begin{table}[]
\centering
\caption{Performance comparison of embedding types.}
\label{tab:embedding_types_comp}
\begin{tabular}{|l|c|c|l|}
\hline
\textbf{Model} & \textbf{Comb.} & \textbf{Description} & \textbf{Direct} \\ 
\hline
all-roberta-large-v1 & $0.71$ & $0.26$ & $0.72$ \\
blade-embed & $0.76$ & $0.66$ & $0.74$ \\
bilingual-embedding-large & $0.62$ & $0.57$ & $0.69$ \\
jina-embeddings-v3  & $0.76$ & $0.46$ & $0.81$ \\
mxbai-embed-large-v1 & $0.71$ & $0.61$ & $0.73$ \\
stella-base-en-v2 & $0.72$ & $0.59$ & $0.68$ \\ 
\hline
\end{tabular}
\end{table}

After this experiment, we evaluate the optimal number of templates to be used to maximize the system performance. We tested scenarios in which a varying number of templates from 1 to 10 were returned.
This analysis revealed that using two templates achieved optimal performance, with an Accuracy and MCC of $0.8$. As shown in Table~\ref{tab:n_templates}, performance declined when fewer or more templates were used.

\begin{table}[]
\centering
\caption{Evaluation of system performance based on the number of templates.}
\label{tab:n_templates}
\begin{tabular}{|l|c|c|}
\hline
\textbf{Model} & \textbf{Accuracy} & \textbf{MCC} \\ \hline
1 Template & $0.70$ & $0.60$ \\
2 Templates & $0.80$ & $0.80$ \\
5 Templates & $0.70$ & $0.60$ \\
7 Templates & $0.60$ & $0.60$ \\
10 Templates & $0.60$ & $0.60$ \\ 
\hline
\end{tabular}
\end{table}

Finally, the confusion matrix in Figures~\ref{fig:template_class_CM} demonstrates the system's performance across query classes. The system achieved near-perfect accuracy for \texttt{SELECT} and \texttt{GROUP BY}, with slightly lower performance for \texttt{FILTER}. 

\begin{figure}[ht]
    \centering
    \includegraphics[width=0.72\columnwidth]{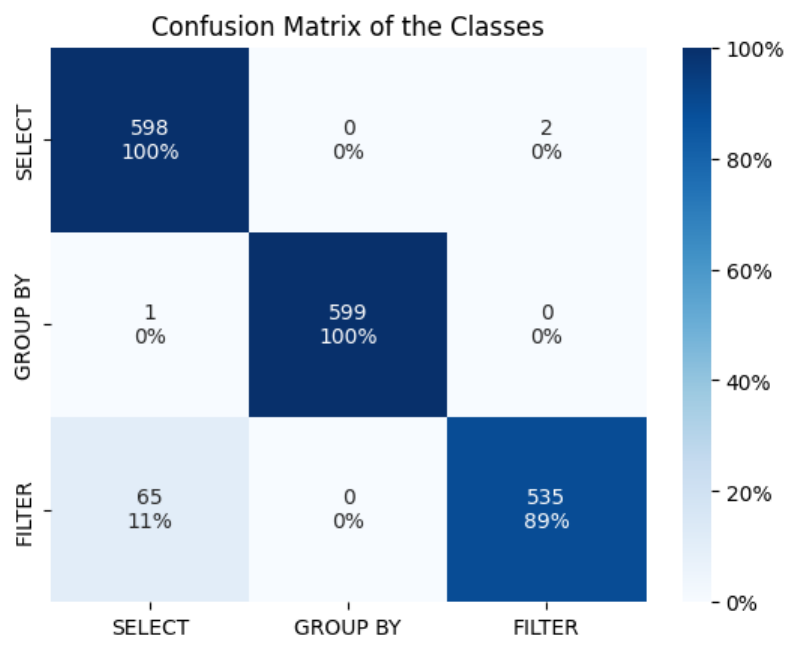}
    \caption{Confusion matrix for query classes.}
    \label{fig:template_class_CM}
\end{figure}

\subsection{Evaluation of the Query Generation Component}

The evaluation of the system's ability to generate SPARQL queries was conducted using a carefully designed dataset and well-defined metrics. This subsection summarizes the findings, including the comparison of large language models (LLMs), an analysis of the best-performing model, and the impact of templates on query generation.

The dataset used in the evaluation was manually constructed to ensure it was both representative and reliable. Each sample comprised a natural language question that represented the user's query, paired with the corresponding SPARQL query reflecting the intended output. Additionally, each sample identified the primary KG entity involved and documented the number of triples retrieved as a result of executing the query on a test dataset. This structured approach ensured the dataset provided a comprehensive foundation for evaluating system performance. For simplicity, in this experiment, the dataset included 24 samples for each of the following four entities extracted from the underlying ontology: \textit{Observation}, \textit{Sensor}, \textit{Observable Property}, and \textit{Platform}. The complexity of each query was classified into one of three levels: \textit{Simple}, \textit{Medium}, or \textit{Complex}. In particular, \textit{Simple} means that a basic attribute retrieval using straightforward \texttt{SELECT} queries is performed; \textit{Medium} means that a filtered retrieval using \texttt{FILTER} clauses is conducted; and finally, \textit{Complex} means that an advanced queries involving \texttt{GROUP BY} operations and filters are executed.

Four metrics were employed to evaluate the system's performance, each designed to assess a specific aspect of query generation and execution. These metrics, along with their corresponding formulas, are detailed below:

\begin{itemize}
    \item \textbf{Execution Success Rate (ESR)} is a metric that measures the percentage of queries executed without errors, providing insight into the system's robustness in generating valid SPARQL queries. It is defined as
    $ESR = \frac{N_{\text{success}}}{N_{\text{total}}}$, where \(N_{\text{success}}\) represents the number of queries executed successfully, and \(N_{\text{total}}\) is the total number of queries submitted for execution.

    \item \textbf{Result Count Accuracy (RCA)} evaluates the accuracy of the number of results retrieved by the query, assessing whether the generated queries return the expected quantity of data. It is given by $RCA = \frac{N_{\text{correct results}}}{N_{\text{success}}}$. Here, \(N_{\text{correct results}}\) refers to the number of successfully executed queries that returned the correct number of results.

    \item \textbf{Result Match Rate (RMR)} assesses the precision of the retrieved data by comparing it to the expected results. It is calculated as $RMR = \frac{N_{\text{matching results}}}{N_{\text{success}}}$, where \(N_{\text{matching results}}\) denotes the number of successfully executed queries that returned data matching the expected content.

    \item \textbf{Harmonic Result Accuracy (HRA)} combines ESR and RMR into a single measure to evaluate both query execution reliability and result precision. By using the harmonic mean, it penalizes scenarios where one of these components is significantly lower. The formula is $HRA = 2 \cdot \frac{ESR \cdot RMR}{ESR + RMR}$.
    
\end{itemize}

Together, these metrics provide a comprehensive evaluation of the system, addressing both its ability to generate executable queries and its effectiveness in retrieving accurate and precise results.

In the first experiment related to this component, we compared several Large Language Models across the above-defined metrics. Figure~\ref{fig:llm_comparison} illustrates the performance of four LLMs: LLAMA $3.1$ 70B, LLAMA $3.1$ 8B, LLAMA $3.2$ 3B, and Qwen 2 72B. The Qwen 2 72B model achieved the highest ESR ($100\%$), indicating robust query execution. However, its RCA ($84.4\%$) and RMR ($66.7\%$) revealed room for improvement in accuracy. In contrast, LLAMA $3.1$ 70B achieved the highest HRA ($82.7\%$), reflecting balanced performance.

\begin{figure}[ht]
    \centering
    \includegraphics[width=\columnwidth]{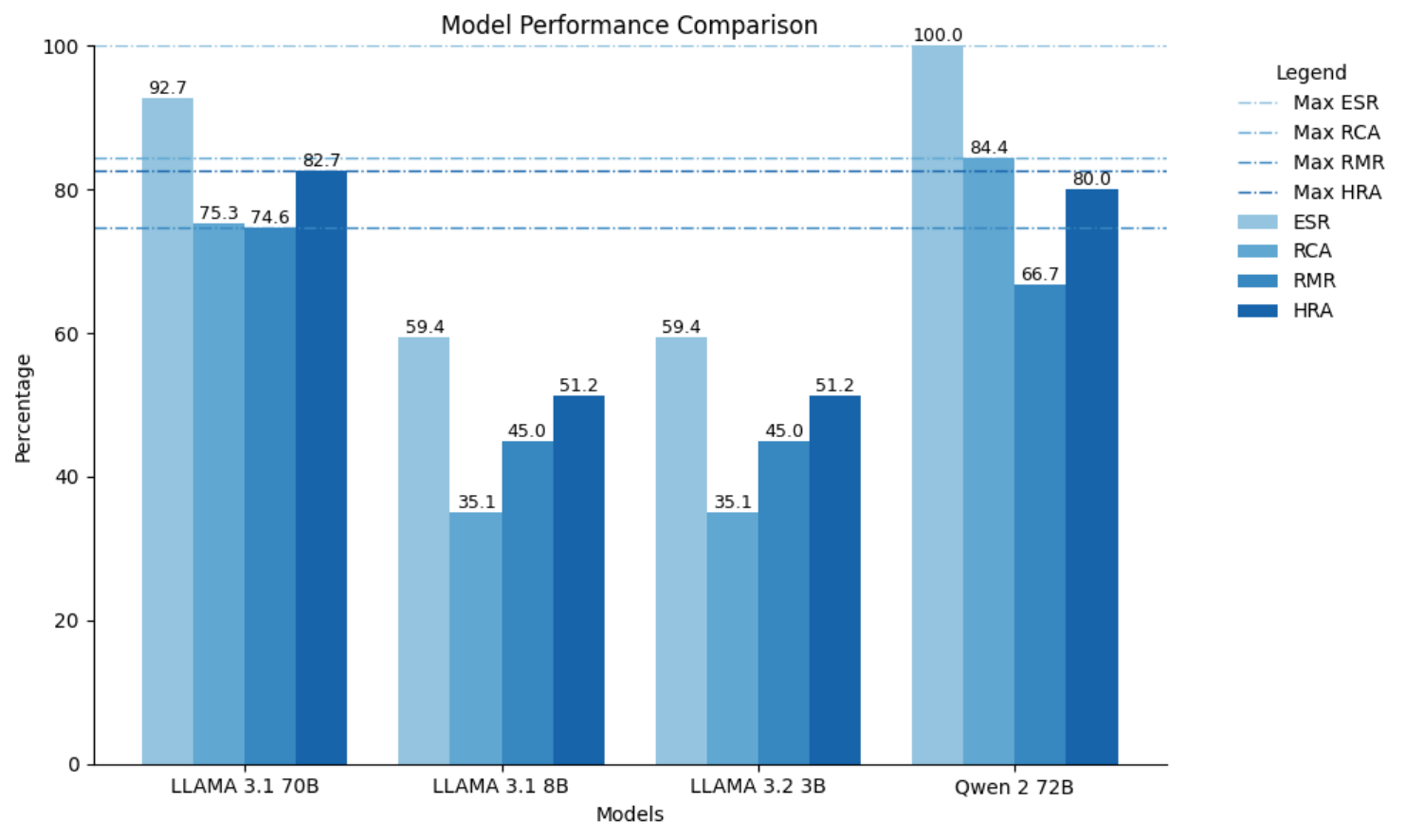}
    \caption{Performance comparison of LLMs across defined metrics.}\label{fig:llm_comparison}
\end{figure}

In the second analysis, we investigated the impact of the presence of the templates.
The use of templates significantly improved performance across all metrics (Figure~\ref{fig:with_without_templates_comp}). Templates provided structured guidance for query generation, enhancing both execution success and result accuracy. Without templates, ESR decreased to $79.2\%$, and RCA and RMR fell to $65.8\%$ and $55.3\%$, respectively.

\begin{figure}[ht]
    \centering
    \includegraphics[width=\columnwidth]{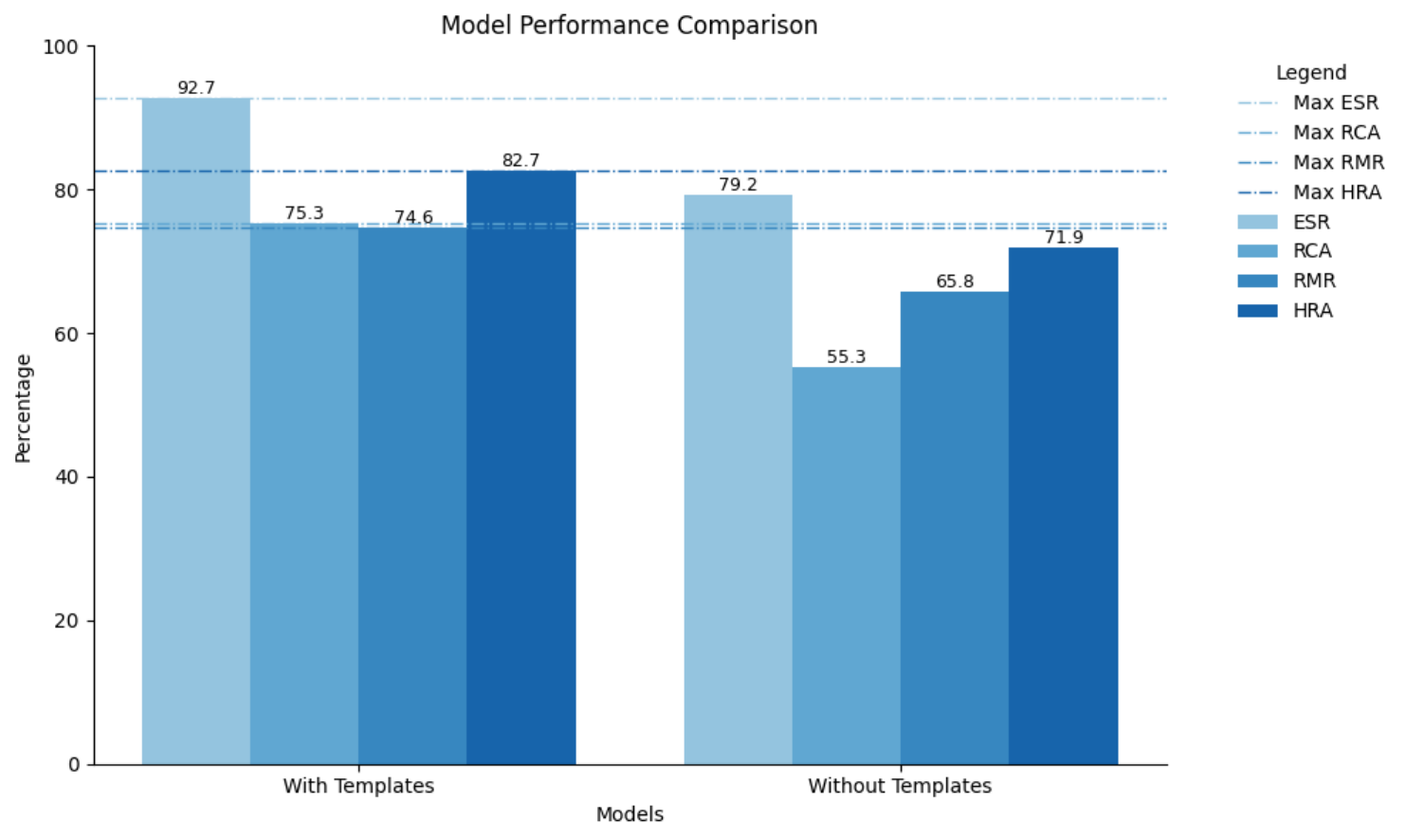}
    \caption{Performance comparison with and without templates.}\label{fig:with_without_templates_comp}
\end{figure}

Finally, we studied the performance according to the above-defined metrics in relation with the four entities of the underlying ontology mentioned above.
The LLAMA $3.1$ 70B model demonstrated strong performance across entities and query types. It achieved a perfect ESR ($100\%$) for queries related to \textit{Observation} and high scores for \textit{Platform} queries ($95.8\%$). However, as shown in Table~\ref{tab:esr_vs_entity}, it exhibited challenges with queries related to the \textit{Sensor} entity. This can be explained by the higher complexity in the relationships modeled by the SOSA/SSN ontology for this entity.

\begin{table}[]
\centering
\caption{ESR performance across KG entities.}\label{tab:esr_vs_entity}
\begin{tabular}{|c|c|c|c|}
\hline
\textbf{\begin{tabular}[c]{@{}c@{}}Observable\\ Property\end{tabular}} & \textbf{Observation} & \textbf{Platform} & \textbf{Sensor} \\ 
\hline
$87.5\%$ & $100\%$ & $95.8\%$ & $87.5\%$ \\ 
\hline
\end{tabular}
\end{table}

\subsection{Evaluation of the Data Visualization Component}

This subsection presents the evaluation of the dashboard generation process, focusing on the system's capability to determine appropriate visualization types and to generate accurate visual representations based on SPARQL query results. The assessment aimed to evaluate the effectiveness and reliability of the system in generating suitable visualizations that align with the underlying data.

The dataset used to evaluate dashboard generation was derived from the query evaluation dataset. However, samples where SPARQL queries failed to execute successfully or did not return valid data were excluded, as visualizations are generated only when valid query results are available. Based on this, the dataset was composed of a total of $67$ samples. Each sample was manually categorized as either ``Plot'' or ``Table'', based on the type of visualization that could better represent the query results. This categorization provided a basis for assessing the system's ability to select the correct visualization type. For the evaluation, the used metric is the accuracy defined before.

The first experiment evaluated the system's ability to select the appropriate visualization type. This result was evaluated by comparing its predictions against the manually labeled dataset. The system accurately identified the correct visualization type in $70\%$ of the samples. These results highlight the system's capability to make informed decisions, although there is room for improvement. To analyze prediction errors, a confusion matrix schematized in Table~\ref{tab:conf_matrix_dashboard} was generated. As shown in this table, the breakdown of correct and incorrect predictions for the ``Plot'' and ``Table'' visualization types is present. The system demonstrated higher accuracy for ``Plot'' predictions ($83\%$) than for ``Table'' predictions ($70\%$). This indicates a stronger alignment with graphic-related visualizations but suggests challenges in distinguishing cases where tabular representation is more suitable.

\begin{table}[]
\centering
\caption{Confusion matrix showing the system's performance in predicting ``Plot'' and ``Table'' visualization types.}\label{tab:conf_matrix_dashboard}
\begin{tabular}{|c|c|c|}
\hline
\textbf{} & \textbf{Plot} & \textbf{Table} \\ \hline
\textbf{Plot} & $83\%$ & $17\%$ \\ \hline
\textbf{Table} & $30\%$ & $70\%$ \\ \hline
\end{tabular}
\end{table}

The second and last experiment assessed the system's ability to generate visualizations accurately. This was measured by calculating the percentage of cases where the visualizations selected by the system were successfully generated. The results revealed that the system generated all selected visualizations without errors, achieving a $100\%$ success rate. This outcome underscores the robustness of the visualization generation process, including the execution of the underlying Python code. The system's iterative refinement approach ensures that visualizations are not only accurately generated but also appropriately tailored to the data characteristics. The high success rate in generating visualizations demonstrates the system's reliability in handling diverse SPARQL query results. It provides users with a seamless and error-free experience, confirming the practicality of the dashboard generation process for real-world applications.

\section{Conclusion}
\label{sec:conclusion}
In this work, a comprehensive system for question answering over KGs is presented, addressing challenges in query generation, and data visualization. The solution starts from the construction of a KG, using a flexible ETL pipeline, ensuring compatibility with the SemIoE ontology. A hybrid SPARQL query generation approach, based on a Retrieval-Augmented Generation (RAG) framework, integrates template matching with Large Language Models (LLMs). Retrieved templates provide context to the LLM, enabling the translation of natural language queries into accurate and interpretable SPARQL queries. This approach balances the adaptability of LLMs with the reliability of predefined templates, mitigating issues such as query hallucination and semantic misalignment. Furthermore, the implementation of a dynamic dashboard facilitates intuitive and interactive visualization of query results. This dashboard automates the selection of visualization types, enabling effective exploration of data while ensuring the accurate generation of graphs and tables. As an additional contribution, we introduced solution-specific evaluation metrics to thoroughly validate the system's performance, thus demonstrating robustness across template retrieval, query generation, and dashboard visualization. These contributions collectively enhance the usability of KGs, empowering users to interact with complex semantic data effectively.

While the system achieved its primary objectives, several areas offer opportunities for improvement. The dashboard's decision-making process for visualization selection could benefit from integrating advanced techniques and supporting a broader range of visualization types, such as heatmaps, geospatial maps, or network graphs. Expanding these options would better accommodate diverse data structures and analytical needs. Incorporating interactive features, such as user-defined preferences for chart types, color schemes, or data groupings, could further tailor visualizations to individual requirements, fostering a more user-centered experience. Real-time query execution and visualization capabilities represent another avenue for future development, addressing time-sensitive applications like IoT monitoring and financial analytics.



\bibliographystyle{plain}
\bibliography{biblio}

\end{document}